\documentclass[11pt]{article}
\usepackage{graphicx}

\newcommand{\BABARPubYear}    {01}

\newcommand{\BABARConfNumber} {71}
\newcommand{\SLACPubNumber} {9045}

\input pubboard/babarsym

\long\def\inst#1{\par\nobreak\kern 4pt\nobreak
    {\it #1}\par\vskip 10pt plus 3pt minus 3pt}

\begin{document}
{\pagestyle{empty}

\begin{flushright}
\babar-CONF-\BABARPubYear/\BABARConfNumber \\
SLAC-PUB-\SLACPubNumber \\
October, 2001 \\
\end{flushright}

\par\vskip 5cm

\begin{center}
\Large \bf Charmless hadronic {\boldmath $B$\unboldmath} decays at \babar\
\end{center}
\bigskip

\begin{center}
Marcella Bona\\
INFN and University of Torino\\
Via Pietro Giuria 1, Torino\\
E-mail: bona@to.infn.it\\
(on behalf of the \babar\ Collaboration)
\end{center}
\bigskip \bigskip

\begin{center}
\large \bf Abstract
\end{center}

Using 22.7M $B\bar B$ events collected with the \babar\ detector at SLAC,
we present preliminary measurements of the branching fractions for charmless hadronic
decays of $B$ mesons into two-body, quasi two-body and three-body final states with
pions, kaons, and $\rho$ and $a_0$ resonances.
In the search for exclusive $B^0 \rightarrow \pi^+\pi^-\pi^0$, we measure
${\cal B}(B^0 \rightarrow \rho^{\pm}(770)\pi^{\mp})=(28.9\pm5.4\pm4.3) \times 10^{-6}$,
together with the relative asymmetry ${\cal A}_{\rm phys} = -0.04 \pm 0.18 \pm 0.02$.
We also set the upper limits on 
$B^0 \rightarrow \rho^0(770)\pi^0$, non-resonant $B^0 \rightarrow \pi^+\pi^-\pi^0$,
$B^0 \rightarrow a_0^{\pm}(\rightarrow \eta\pi^{\pm})\pi^{\mp}$ and
$B^0 \rightarrow K^0\bar K^0$.

\vfill
\begin{center}
Contributed to the Proceedings of the International Europhysics Conference on High Energy Physics,\\
7/12/2001-7/18/2001, Budapest, Hungary
\end{center}

\vspace{1.0cm}
\begin{center}
{\em Stanford Linear Accelerator Center, Stanford University, 
Stanford, CA 94309} \\ \vspace{0.1cm}\hrule\vspace{0.1cm}
Work supported in part by Department of Energy contract DE-AC03-76SF00515.
\end{center}

\newpage

\section{Introduction}
\label{sec:Introduction}
Measurement of the rates and $C\!P$ asymmetries for $B$ decays into the charmless final
states can be used to constrain the angle $\alpha$ of the unitarity triangle \cite{bbphys}.
In the case of three body $\pi^+\pi^-\pi^0$ decays, such measurements of
$\alpha$ would exploit interference between the $B^0 \rightarrow \rho^{\pm}\pi^{\mp}$
modes and the colour-suppressed $B^0 \rightarrow \rho^0\pi^0$.

In the case of $B^0 \rightarrow a_0^{\pm}\pi^{\mp}$, the absence of second-class
currents, together with the assumption of factorization, provide new constraints on $C\!P$
observables. The kinematics do not allow interference between the oppositely-charged
resonances in the Dalitz plot as in the $B^0 \rightarrow \rho(770)\pi$, but in
the absence of second-class currents might lead to enhanced direct $C\!P$ violation  \cite{a0pi}.

In the case of $B^0 \rightarrow K^0\bar K^0$, the decay rate is expected to
be small ($10^{-6}-10^{-7}$) in the Standard Model \cite{k0k0}. Final-state rescattering
effects can lead to enhancement of the branching fraction and the possibility of large
strong phases, with correspondingly large $C\!P$-violating charge asymmetries
\cite{k0k0bis,k0k0ter}.
Observation of the $K^0\bar K^0$ decay mode would provide important information about
the strength of final-state rescattering in charmless $B$ decays.

\section{Analysis}
\label{sec:Analysis}
The data sample used consists of $22.74$ million $B \bar B$ events, collected at the
PEP-II asymmetric $e^+e^-$ collider at SLAC, with the \babar\ detector \cite{det}.
Hadronic events are selected based on track multiplicity and event topology. We use only
good quality tracks.
Candidate $K_S^0$ mesons are reconstructed from pairs of oppositely-charged
tracks that form a well-measured vertex and have invariant mass within $3.5$
standard deviations ($\sigma$) of the nominal $K_S^0$ mass \cite{pdg}.
Candidate photons are defined as showers in the electromagnetic calorimeter that have the
expected lateral shape and are not matched to a track.
Candidate $\pi^0$ mesons are reconstructed by combining pairs of photons with an invariant
mass between $100$ and $160\mevcc$; the $\pi^0$ candidates are then
kinematically fitted with their mass constrained to the nominal $\pi^0$ mass \cite{pdg}. 
Pion candidates (except $K_S^0$ and $a_0$ daughters) are required to fail kaon selection criteria.

We reconstruct the decay $B \rightarrow a_0\pi$ in the mode $a_0\rightarrow \eta\pi$,
$\eta \rightarrow \gamma\gamma$. To be associated with an $\eta$ decay a pair of candidate
photons is required to satisfy $0.470 < m_{\gamma\gamma} < 0.615\gevcc$ and
the $\eta$ center-of-mass (CM) momentum must be larger than $0.9\gevc$.
The pion track and $\eta$ candidate form an $a_0$ candidate if $0.90 < m_{\eta\pi} < 1.08\gevcc$.

Candidate $B$ mesons are selected by exploiting the kinematic constraints provided by the
$\Upsilon(4S)$ initial state. First we define an energy-substituted mass $m_{ES}$,
where $\sqrt{s}/2$ is substituted for candidate's energy \footnote{In $B^0 \rightarrow a_0\pi$
analysis $m_{ES}$ is replaced with the energy-constraint mass $m_{EC} = \sqrt{s/4-p_B^2}$
where $p_B$ is obtained by applying kinematic constraints to the four-momenta of the $B$ daughters.}.
The second variable used is the difference $\Delta E$ between the $B$-candidate
energy and $\sqrt{s}/2$. For all modes the $m_{ES}$ resolution is dominated by the beam
energy spread and is approximately $2.5\mevcc$, while $\Delta E$ resolution is mode dependent
and dominated by momentum resolution.
Candidates are selected in the range $5.2 < m_{ES}(m_{EC}) < 5.3\gevcc$ and
accepted, depending on the decay topology, in various $\Delta E$ ranges,
restrictive enough to suppress background due to other types of $B$ decays.

The largest source of background is from random combinations of tracks and neutrals
produced in the $e^+e^- \rightarrow q\bar q$ continuum (where $q=u, d, s$ or $c$).
In the CM frame this background typically exhibits a two-jet structure. In contrast,
the low momentum and pseudo-scalar nature of $B$ mesons from $\Upsilon(4S)$ decays
leads to a more spherically symmetric event.
This topology difference is exploited using event-shape quantities. The first variable is the
angle $\theta_T$ between the thrust axes, in the CM frame, of the
$B$ candidate and the remaining tracks and photons in the event (ROE).
We require $|\cos\theta_T| < 0.9$.
Another quantity used is a Fisher discriminant $\cal F$, a linear combination of several
discriminating variables like the scalar sum of the momenta of the ROE flowing into nine
concentric cones centered on the thrust axis of the $B$ candidate, in the CM frame~\cite{fisher}.
Another set of discriminating variables is defined by
$L_j^{(c,n)} = \sum_{i_{(c,n)}}p_i\times |cos\theta_i|^2$,
which are the momentum-weighted sums of the cosines of the angles between the
ROE charged tracks ($L_j^{(c)}$) or neutral clusters $L_j^{(n)}$
and the thrust axis of the \B\ candidate. In the analysis of $B \rightarrow a_0\pi$ 
these variables are used in a non-linear (Neural Network) multi-variate analysis.

Global detection efficiencies, including branching fractions of intermediate states,
are listed in Table \ref{table1}. Appropriate control samples are used to determine
efficiencies for $\pi^0$ and $K_S^0$ reconstruction, particle identification,
and selection criteria for $m_{ES}$ and $\Delta E$.

Signal yields are determined with either a simple counting analysis, or with a
maximum likelihood fit.  For the counting analysis, the yield is defined as
$N_S = N_1 - {\cal R}N_2$, where ${\cal R}$ is the background fraction of the
number of candidates in the signal region to the number in the side-band region,
$N_1$ is the number of candidates in the signal region for on-resonance data and
$N_2$ is the number of candidates in on-resonance data observed in the side-band
region.
In the second technique, signal yields are determined from an unbinned maximum
likelihood fit using  $m_{ES}$ or $m_{EC}$, $\Delta E$, ${\cal F}$ or $N\!N$ output,
$\gamma\gamma$ mass (where applicable).
In each of the fits, the likelihood for a given candidate is obtained
by summing the product of event yields and probabilities over all possible signal
and background hypotheses.
Monte Carlo simulated data is used to validate the assumption that the fit variables are
uncorrelated.
The parameters of $m_{ES}$, $m_{EC}$, $\Delta E$ and $\cal F$ PDFs are determined from data
and are cross-checked with Monte Carlo simulation.

\begin{figure}[!htb]
\begin{center}
\includegraphics[width=6.cm]{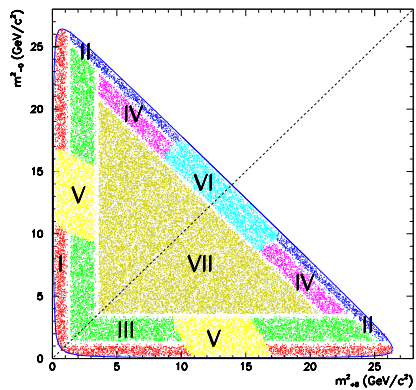}
\caption{Separate region of the Dalitz plot are sensitive to different modes.
	I: $B^0 \rightarrow \rho^{\pm}\pi^{\mp}$. II: $B^0 \rightarrow \rho^0\pi^0$.
	III: $B^0 \rightarrow \rho\prime^{\pm}\pi^{\mp}$. IV: $B^0 \rightarrow \rho\prime^0\pi^0$.
	V: $B^0 \rightarrow$ charged scalar and $\pi^{\mp}$.
	VI: $B^0 \rightarrow$ neutral scalar and $\pi^0$.
	VII: $B^0 \rightarrow \pi^+\pi^-\pi^0$ at high mass.
	}%
\label{fig1}
\end{center}
\end{figure}

Data for the $B^0 \rightarrow \pi^+\pi^-\pi^0$ final state can be represented
on a Dalitz plot (see Fig.~\ref{fig1}). We sub-divide the Dalitz plot into distinct regions,
each of which chosen to be sensitive to a single resonance such as the
$\rho(770)$, $\rho(1450)$ and $f_0(400-1200)$. The regions are defined using
the invariant mass of $\pi\pi$-pair combinations  and the pair helicity angle
defined as the angle between the direction of one of the pions and the direction
of the parent $B$ meson candidate computed in the $\pi\pi$-pair rest frame.
A counting method is used in this analysis.

There are four decay rates that are of interest for the decay mode
$\Bz\to \rho^{\pm}\pi^{\mp}$, defined by
$\Gamma_{\rho\pi} = \Gamma(B^0 \rightarrow \rho^+\pi^-)$ and
$\Gamma_{\pi\rho} = \Gamma(B^0 \rightarrow \rho^-\pi^+)$ together
with their $C\!P$ conjugates $\bar \Gamma_{\rho\pi}$ and $\bar \Gamma_{\pi\rho}$.
A non-zero value for the asymmetry, given by:
\begin{equation}
{\cal A}_{\rm phys} = \frac{(\Gamma_{\rho\pi}+\bar \Gamma_{\pi\rho})-
(\bar \Gamma_{\rho\pi}+\Gamma_{\pi\rho})}
{(\Gamma_{\rho\pi}+\bar \Gamma_{\pi\rho})+
(\bar \Gamma_{\rho\pi}+\Gamma_{\pi\rho})}
\label{eq3}
\end{equation}
would signify direct $C\!P$ violation in at least one of the decays.
\footnote{The numerator in Eq. \ref{eq3} is simply the difference of the two direct $C\!P$ violations
$(\Gamma_{\rho\pi}-\bar \Gamma_{\rho\pi})$ and $(\Gamma_{\pi\rho}-\bar \Gamma_{\pi\rho})$}

\section{Results and Systematics}
\begin{figure}[!b]
\begin{center}
\includegraphics[width=5.0cm]{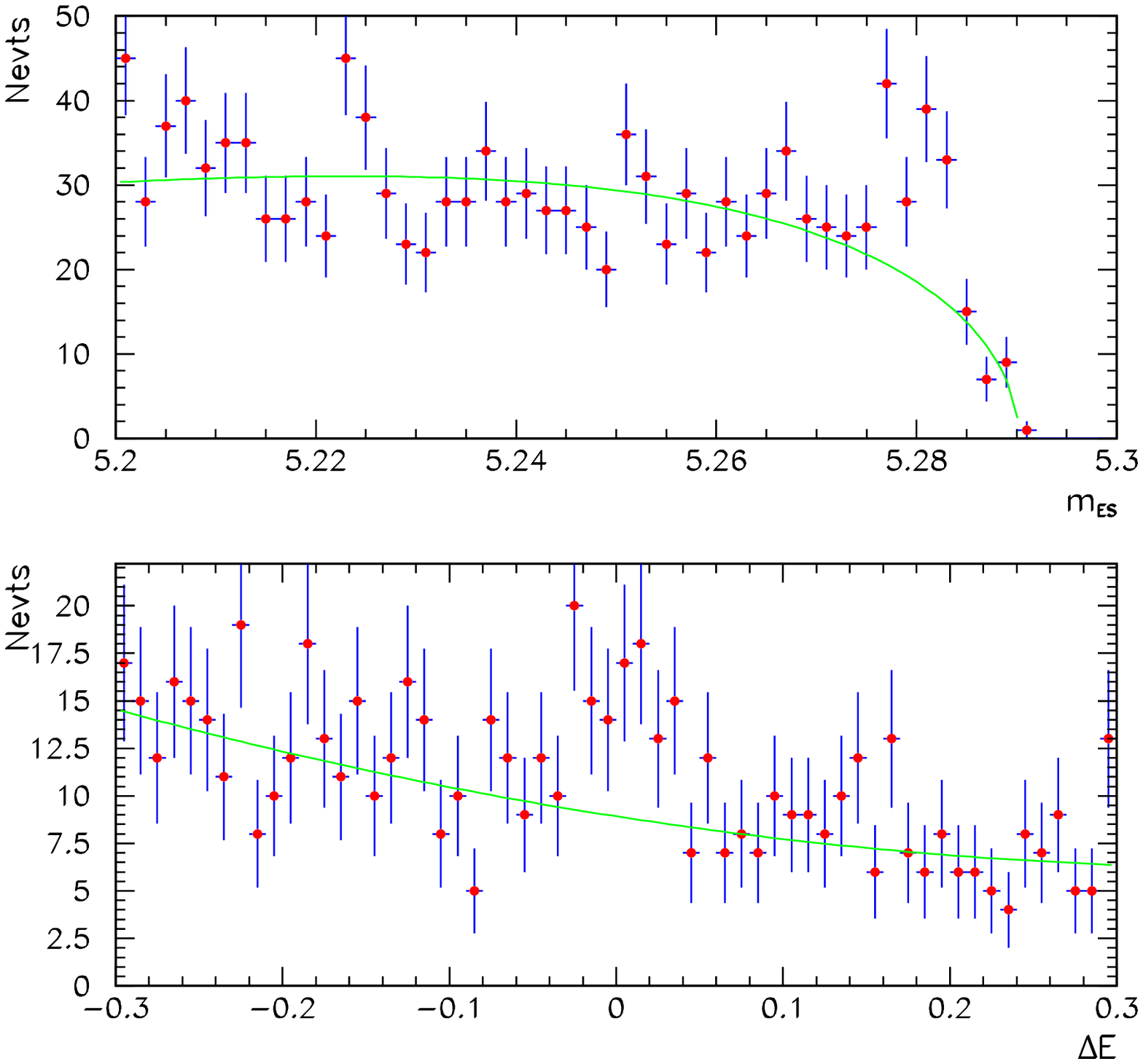}
\includegraphics[width=4.8cm]{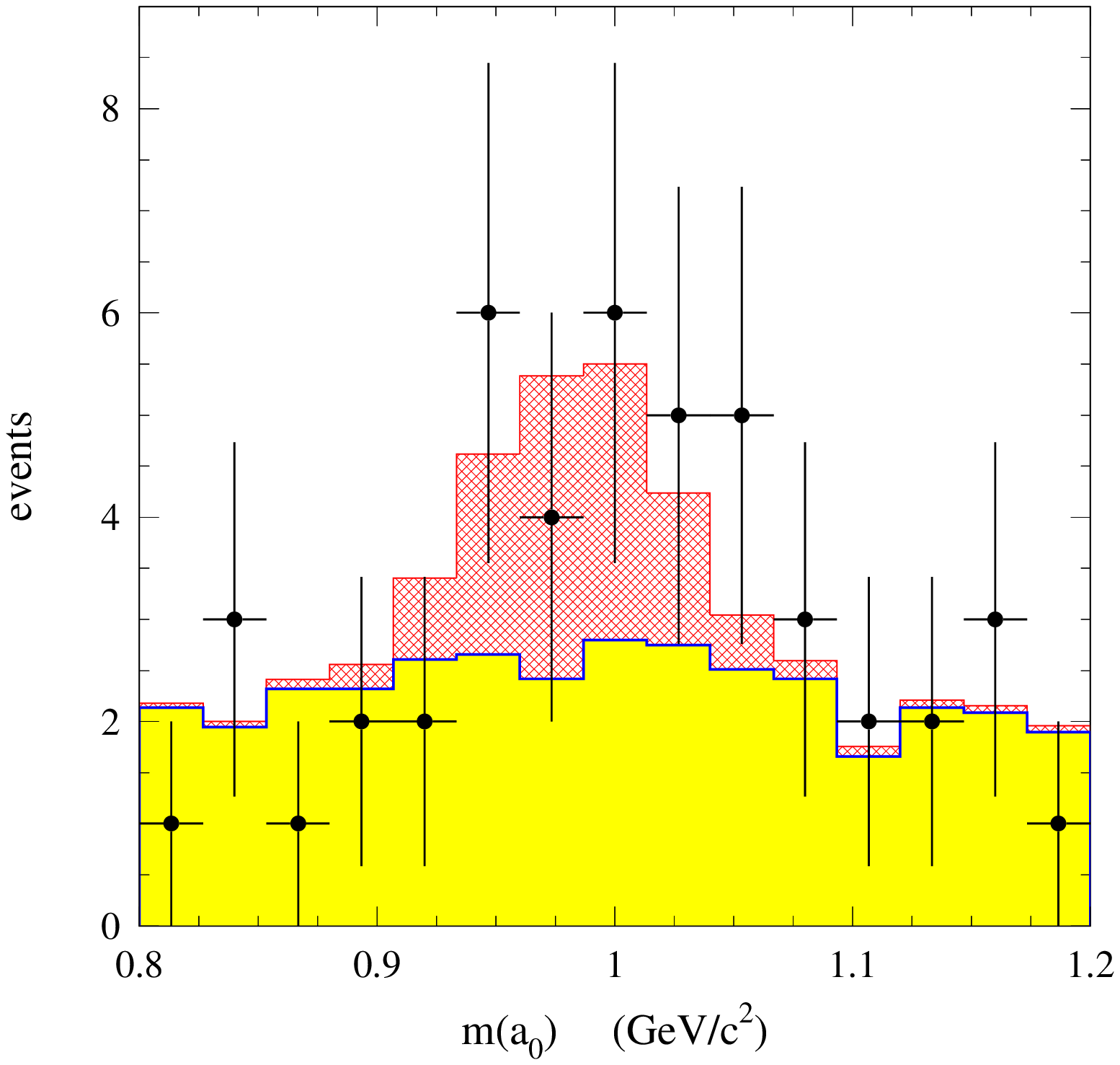}
\includegraphics[width=5.0cm]{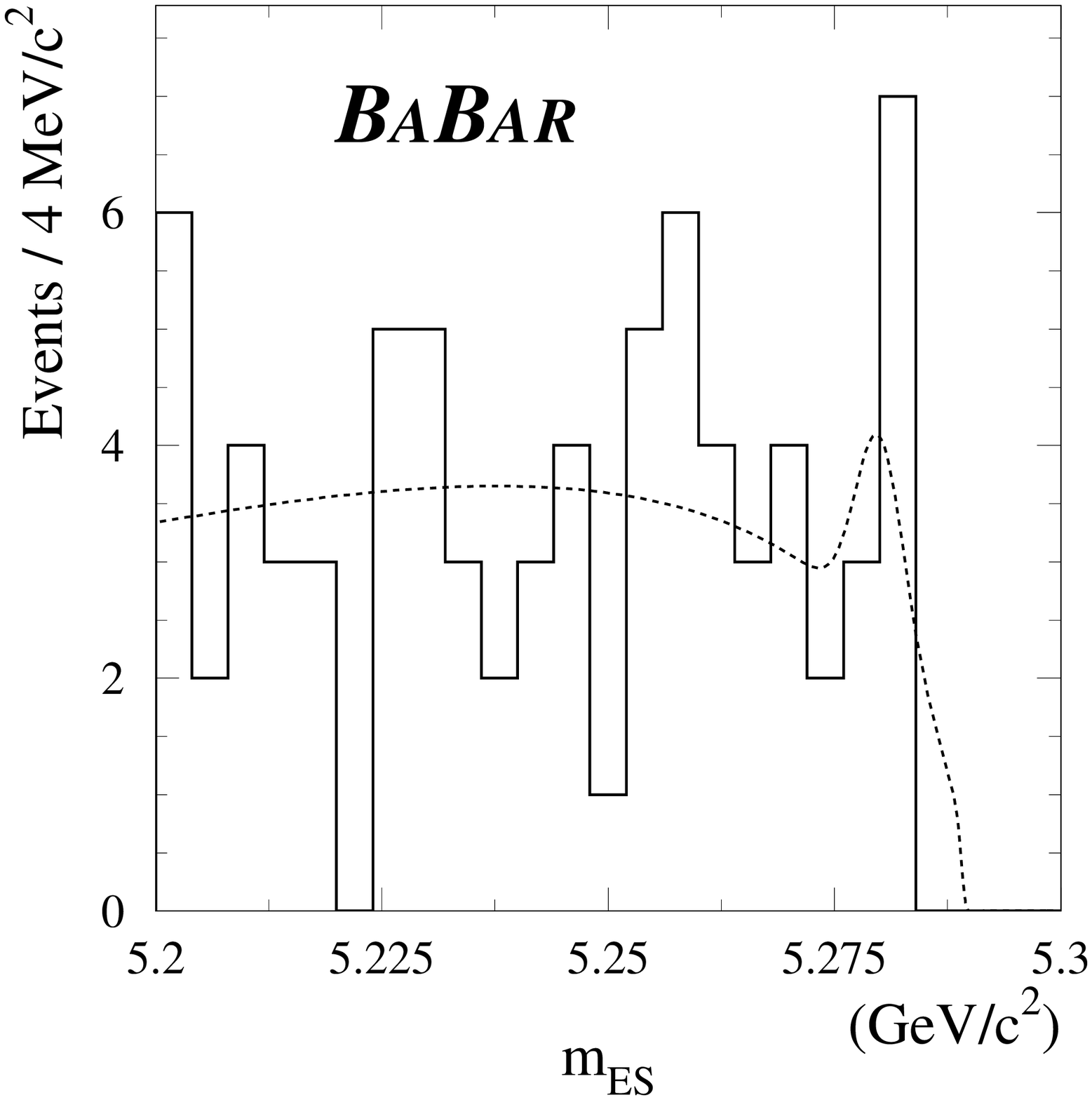}
\caption{Left plot: $m_{ES}$ and $\Delta E$ distributions for $B^0 \rightarrow \rho^{\pm}\pi^{\mp}$.
	Central plot: projection on the $\eta\pi$ invariant mass axis for $B^0 \rightarrow a_0(\eta\pi)\pi$ analysis.
	Right plot: $m_{ES}$ distribution for $B^0 \rightarrow K^0\bar K^0$. The curve is the projection of
	the maximum likelihood fit result. (Projections from likelihood fit are obtained after additional
	requirements on likelihood ratios)}%
\label{fig2}
\end{center}
\end{figure}

The results of the fits or the counting method for the various to\-po\-lo\-gies are
summarized in Table \ref{table1}. In those cases where no evidence of signal is found a $90\%$
confidence level upper limit is computed. In the case of the counting analysis, we have
used the classical method outlined in \cite{bbphys} and we have reduced the background
estimate and the efficiency by one standard deviation (systematic) before making the calculation.
In the case of the maximum likelihood analysis, the upper limit on the signal yield for mode
$k$ is given by the value of $n_k^0$ for which
$\int_0^{n_k^0}{\cal L} dn_k/\int_0^\infty {\cal L}_{max} dn_k = 0.90$
where ${\cal L}_{max}$ is the likelihood as a function of $n_k$, maximized with
respect to the remaining fit parameters.
The result is then increased by the total systematic error, and the detection efficiency
is reduced by its systematic uncertainty in calculating the branching fraction upper
limit. The statistical significance of a given channel is determined by fixing the yield
to zero, repeating the fit and recording the change in $-2\ln{\cal L}$.

We have made a preliminary measurement of the $C\!P$ asymmetry in Eq.~\ref{eq3} of
${\cal A}_{\rm phys} = -0.04 \pm 0.18 \pm 0.02$, which is consistent with zero.
Imperfect knowledge of the PDF shapes, of the detection efficiencies and of the background
subtraction (counting method) are the main sources of systematic uncertainties on
the branching fraction measurements.
Uncertainties in the PDF parameterizations are estimated either by varying the
PDF parameters within $1\sigma$ of their measured uncertainties or by substituting
alternative PDFs from independent control samples and recording the variations in the
fit results.

\medskip
\begin{table}[!htb]
\caption{Summary of results for detection efficiencies ($\epsilon$), signal yields ($N_S$),
statistical significances and measured branching fractions ($\cal B$). Upper limits are
at $90\%$ CL.}
\begin{center}
\begin{tabular}{|l|l|l|l|l|l|}
\hline
mode                        & $\epsilon (\%)$       & $N_S~\pm$ (stat) $\pm$ (syst) & Stat. Sig. ($\sigma$) & ${\cal B} (10^{-6})$\\
\hline
$\rho^{\pm}(770)\pi^{\mp}$  & $13.5\pm 1.6$         & $89\pm16\pm6$                 & $5.0$                 & $28.9\pm5.4\pm4.3$\\
$\rho^0(770)\pi^0$          & $7.4\pm 0.9$          & $6.1\pm5.8\pm2.8$             & $1.0$                 & $< 10.6$\\
$\pi^+\pi^-\pi^0$(NR)       & $7.5\pm 1.0$          & $-4.2\pm7.3\pm3.8$            & N/A                   & $< 7.3$\\
$a_0(\eta\pi)\pi$           & $32.8\pm 2.4$         & $18.1^{+8.7}_{-7.4}\pm 1.6$   & $3.7$                 & $< 11.5$\\
$K^0\bar K^0$               & $36.6\pm 4.6$         & $3.4^{+3.4}_{-2.4}\pm3.5$     & $1.5$                 & $< 7.3$\\
\hline
\end{tabular}
\end{center}
\label{table1}
\end{table}

\section{Summary}
\label{sec:Summary}
We have measured branching fractions for the rare charmless decay
$B^0 \rightarrow \rho^{\pm}(770)\pi^{\mp}$ with its asymmetry
${\cal A}_{\rm phys}$ and set upper limits on $B^0 \rightarrow \rho^0(770)\pi^0$,
non-resonant $B^0\rightarrow \pi^+\pi^-\pi^0$,
$B^0 \rightarrow a_0^{\pm}(\rightarrow \eta\pi^{\pm})\pi^{\mp}$ and
$B^0 \rightarrow K^0\bar K^0$.

\section{Acknowledgments}
\label{sec:Acknowledgments}
We are grateful for the 
extraordinary contributions of our \pep2\ colleagues in
achieving the excellent luminosity and machine conditions
that have made this work possible.
The collaborating institutions wish to thank 
SLAC for its support and the kind hospitality extended to them. 
This work is supported by the
US Department of Energy
and National Science Foundation, the
Natural Sciences and Engineering Research Council (Canada),
Institute of High Energy Physics (China), the
Commissariat \`a l'Energie Atomique and
Institut National de Physique Nucl\'eaire et de Physique des Particules
(France), the
Bundesministerium f\"ur Bildung und Forschung
(Germany), the
Istituto Nazionale di Fisica Nucleare (Italy),
the Research Council of Norway, the
Ministry of Science and Technology of the Russian Federation, and the
Particle Physics and Astronomy Research Council (United Kingdom). 
Individuals have received support from the Swiss 
National Science Foundation, the A. P. Sloan Foundation, 
the Research Corporation,
and the Alexander von Humboldt Foundation.

\end{document}